\documentclass[aps,preprint,showpacs]{revtex4-1}

\usepackage{amsmath,amssymb,bm,graphicx,epsfig,psfrag,ulem,color,natbib}
\usepackage{units}
\usepackage{float} 
\usepackage{amsmath}   
\usepackage{upgreek}  
\usepackage{bm}  
\usepackage[latin1]{inputenc}
\usepackage{epstopdf}
\bibpunct{}{}{,}{s}{}{,}

\begin{document}

\title{Investigation of the momentum distribution of an excited BEC by free expansion: coupling with collective modes}

\author{ A. Bahrami, P.E.S Tavares , A.R. Fritsch, Y.R. Tonin, G.D. Telles, V.S. Bagnato, E.A.L. Henn}

\affiliation{Instituto de F\'isica de S\~ao Carlos, Universidade de S\~ao Paulo, Caixa Postal 369, 13560-970 \\S\~ao Carlos, SP, Brazil}
\email{ehenn@ifsc.usp.br}

\begin{abstract}

We investigate the evolution of the momentum distribution of a Bose-Einstein condensate subjected to an external small oscillatory perturbation as a function of the in-trap evolution of the condensate after the external perturbation is switched-off. Besides changing its momentum distribution, we observe that the cloud distributes the input energy among its normal collective modes, displaying center-of-mass dipolar mode and quadrupolar mode. While the dipolar mode can be easily disregarded, we show that the momentum distribution is closely tied to the quadrupolar oscillation mode. This convolution hinders the actual momentum distribution.
\end{abstract}

\maketitle

\section{Introduction}

Quantum ultracold clouds are object of regular studies since almost 20 years now. The basic tool for revealing their properties is the observation of their atom-density distributions after free expansion. For weakly interacting clouds, this kind of data are direct maps of the in-trap momentum distribution of the clouds \cite{md1}. Free expansion for a strongly interacting sample reveals the conversion of interaction energy into momentum \cite{Castin}.

Among all possible applications for the ability to reveal the momentum distribution, the study of quantum turbulence is one that can better profit. The reason behind it is the fact that turbulence is characterized by a very specific momentum distribution with a well defined power-law behavior, known as the Kolmogorov scaling law \cite{Kolmogorov}. This behavior has been seen in classical turbulence \cite{CT}, quantum turbulence in superfluid liquid Helium \cite{QT} and predicted to take place in quantum turbulence in atomic superfluids \cite{tsubota}. The emergence of Quantum Turbulence \cite{Henn} in a $^{87}$Rb Bose-condensed sample, has created the necessity to obtain the momentum distribution to investigate the similarities and differences with other turbulent systems, either classical or quantum.

In a previous work \cite{thompson}, where we have analyzed the momentum distribution of turbulent clouds, we have observed a linear decrease in the momentum distribution with an exponent that does not match the one expected from a Kolmogorov spectrum, $-\frac{5}{3}$, but is in on the order of -3. Although there are several reasons for this apparent discrepancy, we show here that there might be a more prosaic explanation, not investigated before: the coupling of the analysis of the momentum distribution of the quantum clouds with the their normal modes of oscillation.

In this work, we investigate further the momentum distribution of Bose-Einstein condensates (BEC) subjected to an external perturbation but away from the turbulent regime. We show that the features observed in the momentum distribution extracted from those clouds are coupled to the quadrupolar mode of shape oscillation of the cloud. In this sense, the actual momentum distribution is hindered by the shape oscillation and although we believe the general features are still revealed, the details still lack a method for deconvolution of the cloud momentum distribution and shape oscillation. The correct understanding of this coupling may provide alternatives for the uncoupling and subsequent investigation of the ``pure'' momentum distribution.

\section{Experimental details}

The experimental setup and procedure have been described extensively and in detail in several other prior publications \cite{thompson,vortexformation}. In summary, we produce a $^{87}$Rb BEC containing 2$\times$10$^5$ atoms in a purely magnetic trap with trapping frequencies given by $\omega_r=2\pi\times 188$ Hz and $\omega_z=2\pi\times 21$ Hz. A small sinusoidally time-varying magnetic perturbation is superimposed to the trapping field. The spacial profile of this perturbation is anti-Helmholtz-like with the strong axis making a small angle ($\leq 5^o$) with the trap axis. Similarly, the zero-field point is also slightly shifted from the trap bottom. The frequency of the external field is kept fixed at 189 Hz and the field gradient varies from zero to a maximum value which, throughout this paper, has been varied from 0.08 G/cm up to 0.8 G/cm. The perturbation is left on the trap for 31.7 ms or 6 full cycles. Both frequency and duration of the external field are chosen to the best coupling on the BEC and signal-to-noise ratio of the results without loose of generality.

After the perturbation is switched-off we let the excited cloud evolve by up to 40 ms and we identify this time as t$_{hold}$. We then release the cloud from the trap, letting it expand freely for 20 ms and performing an standard absorption image by a weak probe beam in a CCD camera. All analysis throughout this work have as a starting point the obtained absorption images. 

The momentum distribution of the clouds are extracted with the same procedure performed in our previous work \cite{thompson}. In brief, we assume that the cloud released from the trap expands ballistically and the image after free expansion can be directly mapped in the {\it in-situ} momentum distribution of the cloud. While this is not completely true for strongly interacting clouds, in this work, as before \cite{thompson}, we argue that the interaction energy is small compared to the kinetic energy due of the excited cloud and use the direct mapping. The {\it in situ} momentum distribution is analyzed be performing a polar integration of concentric rings around the center-of-mass of the cloud, which ultimately gives us the dependence of the density as a function of the radius of the cloud. At this point, it is important to emphasize that the obtained momentum distribution, which we call $n'(k')$, represents a 2D momentum distribution, as a consequence that the absorption image is a 2D integrated projection of the actual density distribution. While in our previous work \cite{thompson} we show how can we recover the full momentum distribution through an Abel transformation of $n'(k')$, in this paper we restrict ourselves to the 2D momentum distribution since it is already enough to illustrate the effects we want to describe.

In what follows we plot and analyze $n'(k')$ as a function of t$_{hold}$ and excitation maximum amplitude. In order to simplify the analysis, we normalize the measured atom-number of the obtained data by fixing the integral of $n'(k')$ over the whole $k'$ range to a single value.

\section{Results}

Independent of the excitation amplitude, the overall behavior of the cloud after excitation is as follows: the cloud undergoes center-of-mass dipolar motion together with a scissors-like mode, characterized by the periodic tilting of the cloud axis with respect to the original orientation and a quadrupolar mode, characterized by the simultaneous change of crossed dimensions. For smaller amplitudes of excitation, almost no heat is observed and the amplitude of these modes is small. With increasingly excitation amplitude, the heating of the cloud increases as well as the amplitude of the modes oscillations and we eventually observe the formation of dips in the density distribution which we identify as vortices \cite{vortexformation}. 

\begin{figure} [h!]
\begin{center}
\includegraphics[
  width=0.85\linewidth,
  keepaspectratio]{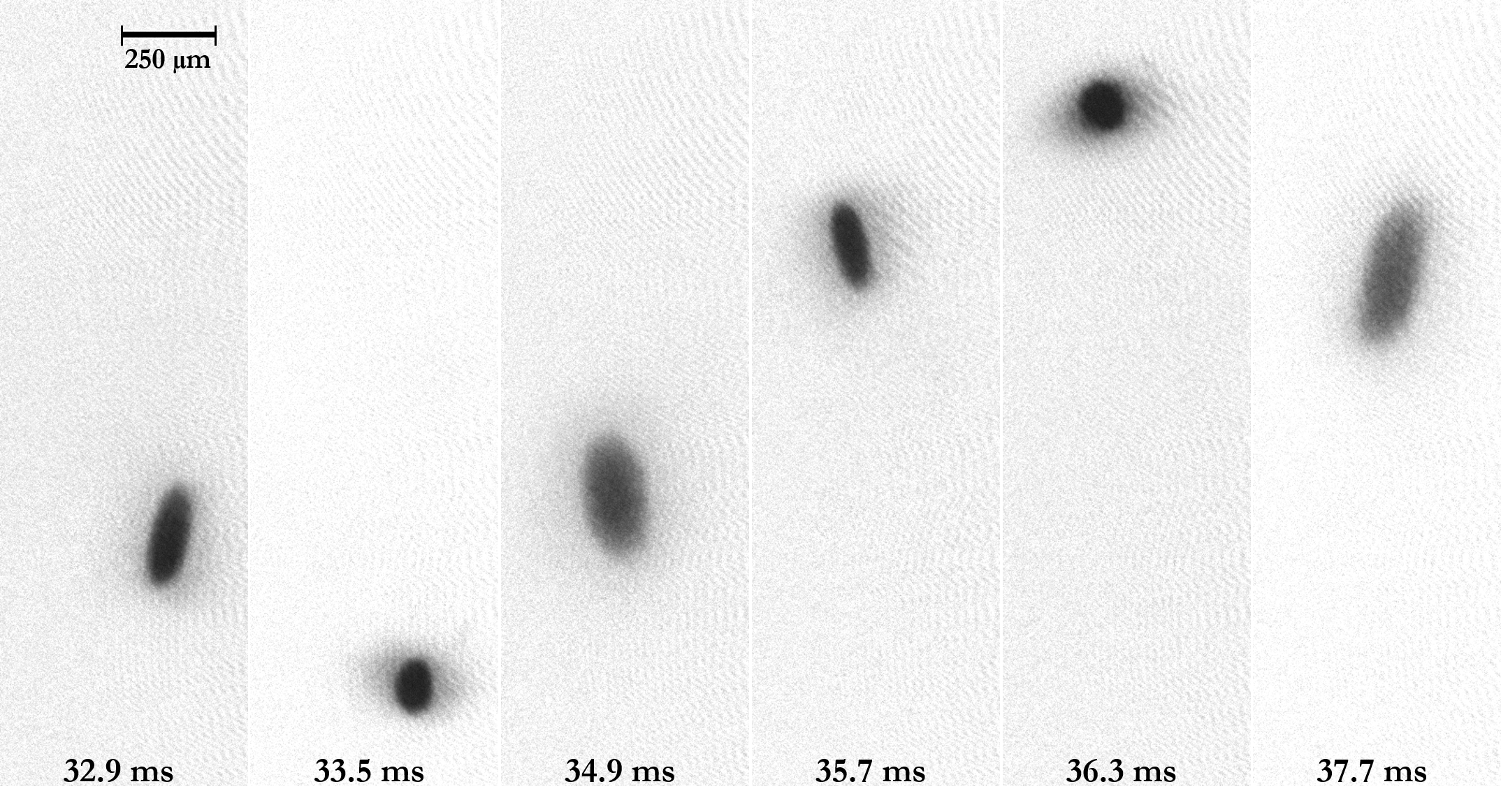}
\end{center}
\caption{Typical data for the cloud behavior after some in-trap t$_{hold}$ evolution time (shown below each picture)  after the excitation (470 mG/cm in this particular sequence) is removed. The behavior is almost independent from the excitation amplitude (see text) only with increasing amplitudes of the oscillation modes, namely the center-of-mass dipolar oscillation, the long-axis periodic tilting attributed to a scissors-like mode and the quadrupole collective shape oscillation.}
\label{evolve}
\end{figure}

The evolution of the cloud and the frequency of the modes can be tracked by observing the cloud at different values of t$_{hold}$ under the same excitation conditions. We show in Fig.\ref{evolve} a typical dataset of images for several t$_{hold}$ equivalent to one oscillation in the trap. Since all the modes are periodic and the characteristic frequencies of the other modes are higher than the dipolar oscillation, for every excitation condition, we follow the cloud for one period of the dipolar oscillation and readily obtain 2 or more periods of the other modes of the cloud.

\begin{figure} [h!]
\begin{center}
\includegraphics[
  width=1.0\linewidth,
  keepaspectratio]{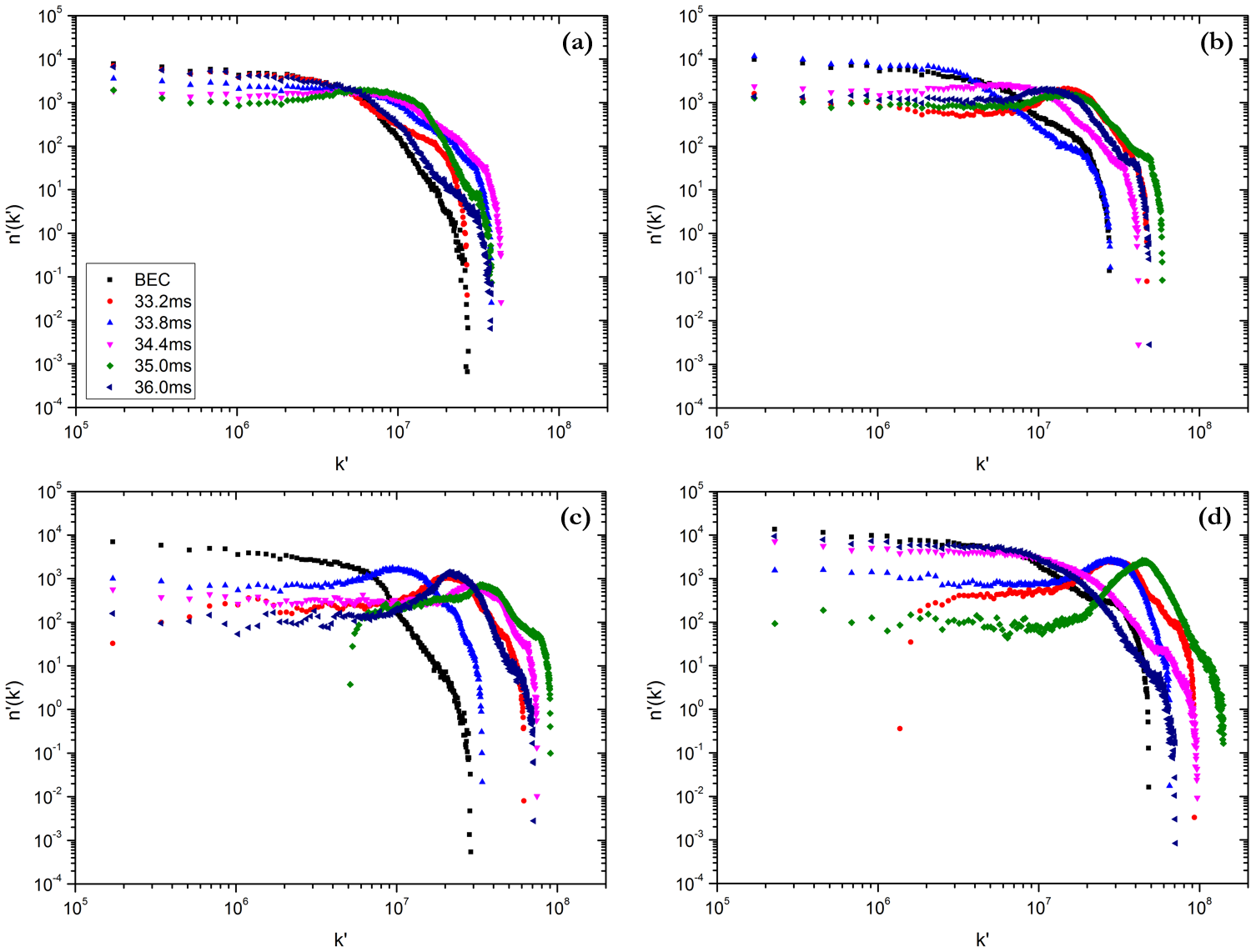}
\end{center}
\caption{(Color online) 2D momentum distribution extracted for different in-trap evolution times t$_{hold}$ for several excitation amplitudes: (a) 315 mG/cm, (b) 470 mG/cm, (c) 630 mG/cm and (d) 710 mG/cm.}
\label{md}
\end{figure}

We analyze each image individually by centering in its center-of-mass and tilting the axis of reference to match the principal axis of the cloud in the vertical axis of the picture. From that simple procedure we immediately get rid of the dipolar motion and the scissors-like mode. We then extract the normalized 2D momentum distribution of the clouds. Fig. \ref{md}(a)-(d) shows the results for several different conditions together with the momentum distribution extracted from a non-excited regular BEC for comparison.

As in our previous work we observe the characteristic momentum distribution of the cold clouds: a plateau for smaller momenta, an almost linear decrease (in log scale) in the mid-region and a sharp decrease in the maximum momentum value of the cloud. Despite the same overall shape of the curves, it is clear that all three regions, namely the plateau, the linear decrease and the final maximum momentum, change their values/slope as a function of the evolution time. 

From our previous work, where we extracted 3D momentum distributions of turbulent clouds at a {\bf fixed} in-trap evolution time, we know that the 3D distribution has a similar shape and the linear decreasing region has an exponent of about $-2.9\pm0.3$, close but not the same of that predicted by the Kolmogorov spectrum for turbulence, $-\frac{5}{3}$. Although there are several reasons for this to occur, like the finite-size of the cloud, the validity of the assumptions on the analysis and even the fact that the Kolmogorov scaling might not be applicable to turbulence in BEC against previous theoretical predictions\cite{tsubota}, the data shown in Fig.\ref{md}(a)-(d) can shine some light on reasons for this discrepancy. It is clear from there that the momentum distribution changes its shape as a function of the in-trap evolution time, an aspect that has not been investigated in previous works. In other words, the excitation of normal modes of the cloud, a collateral effect of our method of excitation, modifies the size and shape of the cloud and affect the momentum distribution. In that sense, the quadrupolar mode hinders the actual momentum distribution by deforming the cloud further than the simple effect of the kinetic energy pumped in the cloud by our excitation.

In order to illustrate the effect of the coupling between the momentum distribution and the quadrupolar mode, we can take any of the characteristics of the curves and analyze it further. For this work, we choose the maximum momenta, where  $n'(k')$ curves fall sharply. 

\begin{figure}
\begin{center}
\includegraphics[
  width=0.85\linewidth,
  keepaspectratio]{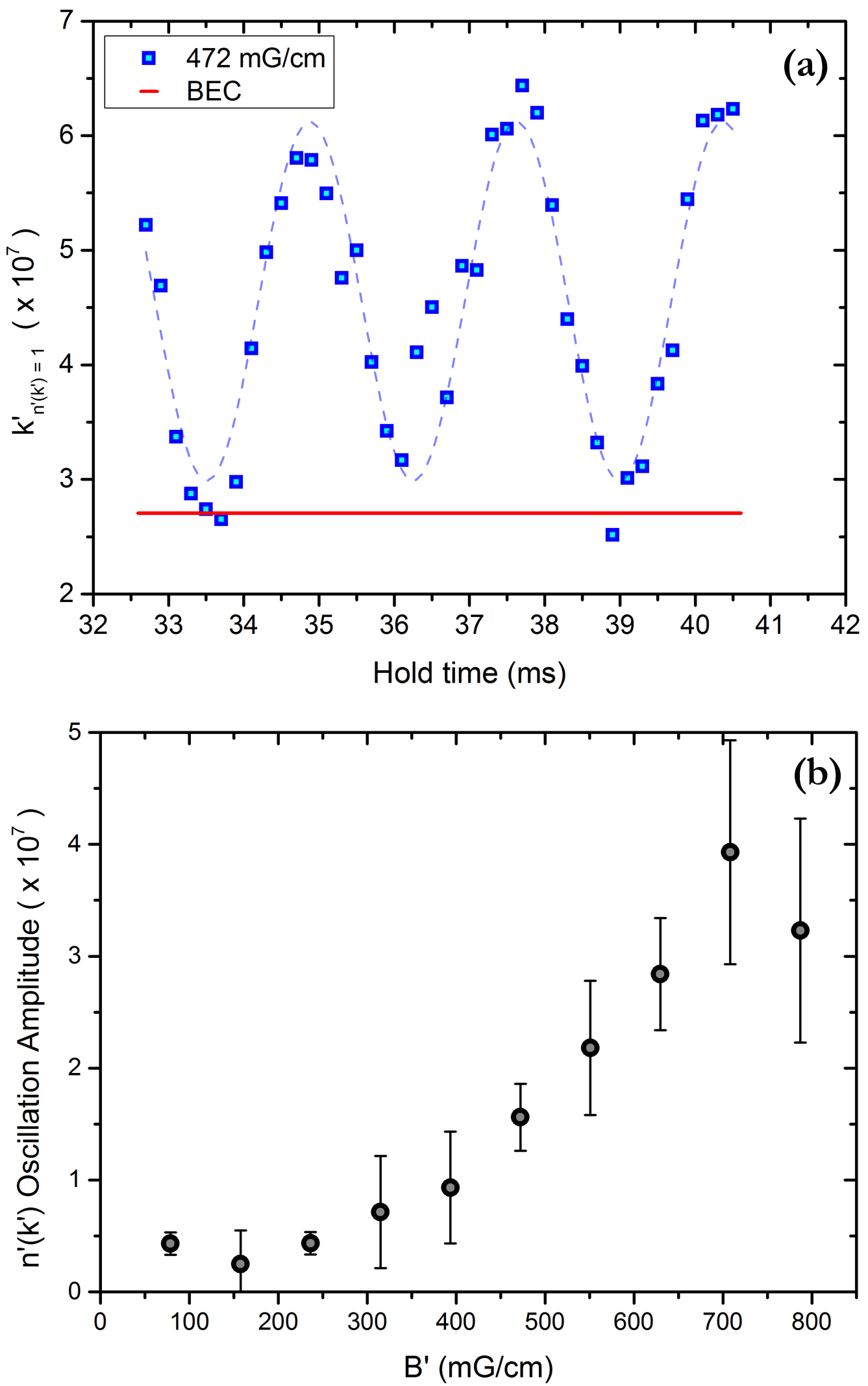}
\end{center}
\caption{(Color online) (a) Oscillation of the maximum momenta of the cloud $k'_{n'(k')=1}$ as a function of evolution time showing a sinusoidal behavior with characteristic frequency matching the quadruple mode frequency of the BEC and (b) amplitude of the fitted oscillation as a function of the excitation amplitude.}
\label{nk}
\end{figure}

We plot in Fig.\ref{nk}(a) the momenta where each of the momentum distribution crosses the level $n'(k')=1$ versus the evolution time for a single excitation amplitude, the one illustrated in Fig.\ref{evolve} and Fig.\ref{md} (b). The behavior could not be more explicit and is the same throughout all the excitation amplitudes studied here: the maximum momenta evolve sinusoidally with a frequency $\omega\approx2\pi\times360$Hz which corresponds to the quadrupolar mode frequency expected for our system given the trapping frequencies of our experiment. The red line gives the reference for a non-excited BEC. Fig.\ref{nk} (b) shows the almost monotonic increase of this amplitude of the oscillation measured as a function of the excitation amplitude, an extra indication of the coupling with the quadrupolar mode. The observed decrease for very high amplitudes is attributed to the system entering non-harmonic regions of the trap and so developing non-linear behavior of its modes of oscillation.

The main consequence of this coupling is that one cannot at this moment extract a precise quantitative analysis of the momentum distribution of the cloud. It is clear that the main qualitative features are present, given the similarities of the curves but one still has to figure out how to decouple the collective mode. The further understanding of this coupling might shine some light on how to decouple them and eventually on the understanding of the momentum distribution of Bose-Einstein condensates under various conditions, in particular displaying vortices and turbulence.

\section{Conclusions}

In summary, we have investigated the dynamical behavior of the 2D momentum distribution of a BEC subjected to an external excitation and we show that the behavior of the extracted momentum distribution is coupled to the quadrupolar normal mode of oscillation of the cloud. The coupling is demonstrated by showing that a characteristic feature of the cloud, namely, the maximum momentum value of the distribution oscillates with a frequency equal to the quadrupolar frequency and the amplitude of this oscillation grows with the oscillation amplitude of the mode which, in its turn, grows with the external oscillation amplitude. The observed effect is a collateral effect of our excitation technique. In order to further study the momentum distribution of such clouds and even applying it to other configurations, like quantum turbulence, we must find a way to deconvolute both effects.

\begin{acknowledgements}
We acknowledge financial support from FAPESP and CNPq.
\end{acknowledgements}

\pagebreak

\end{document}